\documentclass[prl,twocolumn,floatfix,showpacs,amsmath,amssymb] {revtex4}
\usepackage{bm}
\usepackage{graphicx}

\begin{document}

\title{Observing the Inverse melting of the vortex lattice in
  Bi$_2$Sr$_2$CaCu$_2$O$_8$ with point defects using Langevin
  simulations with vortex shaking}

\author{Yadin Y. Goldschmidt and Jin-Tao Liu}
\affiliation{Department of Physics and Astronomy, University of Pittsburgh,
Pittsburgh, Pennsylvania 15260}
\date{\today}
\begin{abstract}
Langevin dynamics simulations of the vortex matter in the
highly-anisotropic high-temperature superconductor
Bi$_2$Sr$_2$CaCu$_2$O$_8$ were performed. We introduced point defects
as a smoothened distribution of a random potential. Both the
electromagnetic and Josephson interactions among pancake vortices were
included. A special shaking and annealing process was introduced to let
the system approach the equilibrium configuration. We are able to see
the inverse melting transition from the Bragg-glass to the
amorphous vortex glass state, in agreement with recent experiments.
\end{abstract}

 
\maketitle
Recent experiments have revealed rich and interesting phase diagram of
the vortex matter in the anisotropic high-temperature superconductor
Bi$_2$Sr$_2$CaCu$_2$O$_8$ (BSCCO) when point impurities are present
\cite{Khaykovich96,Khaykovich97,Fuchs98,Avraham01,Beidenkopf05,Beidenkopf07}. 
Point defects are present even in pristine crystalline samples due to
oxygen vacancies. In addition, one can increase the concentration of
such defects
by bombarding the sample with electrons, the energy of which is of the
order of a 
few MeV \cite{Khaykovich97}. 

In this letter we concentrate on the
effect of point disorder on the location and nature of the first order
(FO) melting line of the vortex lattice with quasi long-range order 
into a vortex liquid or an amorphous vortex glass, depending on the
temperature. 
At high temperatures, in the range of 
$T \gtrsim 0.5 T_c$, where $T_c\approx90K$ for BSCCO, point defects 
have only a mild
effect on the position of the melting line in the $T$-$B$ plane. They
shift the melting line of the vortex solid into a disordered vortex
liquid to a slightly lower temperature for a given magnetic field
\cite{Khaykovich97,Ertas96,Goldschmidt97}. 
The shift becomes larger as the temperature decreases along the
melting line in the $T$-$B$ plane until a special
point is reached with $T=T_{sp}$. Below this temperature the role of
the point disorder becomes much more significant and the glassy
behavior becomes dominant. For $T<T_{sp}$ it is expected that the
melting transition 
is from a dislocation free Bragg glass (BrG) into a dislocation rich
amorphous vortex glass  
(VG) phase which replaces the vortex liquid phase (VL) at high
fields. The melting 
line ceases to be a monotonically decreasing function of $T$ in the
$T$-$B$ plane, and at least near and below $T_{sp}$ it is a
monotonically increasing function of temperature \cite{Beidenkopf05,
Beidenkopf07}. Thus as the temperature is lowered for magnetic fields 
slightly below $B_{sp}$, the field
corresponding to $T_{sp}$, one expects to see first a transition from a
VL to a BrG, which is a solid with quasi-long range order 
\cite{Giamarchi94}, and then a second transition into
a disordered VG phase. The latter transition, which is quite a rare
phenomenon in physical systems, since usually lowering the temperature
results in more order, is referred to as the inverse melting transition
\cite{Avraham01} or sometimes as reentrant behavior
\cite{Ertas96}. Previous theoretical work on the effects of point 
defects on the phase diagram of BSCCO is given in Refs. 
[\onlinecite{Giamarchi94,Ertas96,Goldschmidt97,Kierfeld00,Li06}],
and previous numerical investigations is given in Refs. 
[\onlinecite{Ryu96,Otterlo98,Nonomura00,Olsson00,Olson03,Dasgupta07}].

In this work we observed the inverse melting in numerical
simulations and obtained a phase diagram in qualitative agreement with
experiments. We also investigate the properties of the different
phases near the transition, both statical and dynamical, and
characterize their differences. The method we use for our simulations
is Langevin dynamics, also commonly referred to as Brownian
dynamics. In BSSCO the flux-lines (FLs) which are threads of magnetic
field penetrating the sample and surrounded by revolving currents
(vortices) are more faithfully described by stacks of pancake vortices
\cite{Clem91}, each moving in one CuO$_2$ plane, and interacting with each
other by electromagnetic interaction as well as by the Josephson interaction.
When simulating the dynamics we use the so called over-damped limit
where the mass of the pancakes is negligible and the velocity of each
pancake is proportional to the force acting
on it. This force results from the interaction with other pancakes,
from the effect of the defects, from the thermal white noise, and
possibly from an applied force typically resulting from a Lorentz
force when a global current passes through the sample. 

Some of the advantages of using Langevin 
simulations are that one can control all the interactions precisely, and
one can take microscopic pictures and measure physical quantities that
are difficult to measure in experiments, like the magnitude of the
transverse fluctuations of FLs or the amount of their entanglement. 
In addition it is possible to measure dynamical properties relating to the
pinning and the critical current. It is possible to characterize the
differences between the various thermodynamic phases of the vortex
matter and check in a controlled way how they result from the applied
ingredients and parameters of the model. 
A main disadvantage of the simulations is that the system simulated 
is small and hence phase transitions are always broadened and are not 
as sharp as those observed in real experiments. In our current 
simulations we use 1296 pancake vortices, 36 in each of 36 planes.
The system is a box of size $6a_0\times 3\sqrt{3}a_0\times 36d$ where
$d$ is the CuO$_2$ plane separation,
$a_0=(2\phi_0/\sqrt{3}B)^{1/2}$, $B$ is the magnetic field induction
and $\phi_0$ is the flux quantum.
Pairwise interactions among all pancakes are implemented as discussed 
in detail in
Ref. \onlinecite{Goldschmidt05,Goldschmidt05b,Goldschmidt07}.
Both electromagnetic and Josephson interactions are included. The
former between any pair of pancakes, the 
latter only between nearest neighbor pancakes belonging to the same stack.  
Periodic boundary conditions are used in all directions, including
$z$-direction. We also implement flux cutting and recombination. The
critical temperature for BSCCO was assumed to be $T_c=90$ K. Other
parameters used were $\lambda(0)=1700$ \AA~(penetration depth),
$\xi(0)=30$ \AA~(coherence length), $d=15$ \AA,
$\gamma=375$~(anisotropy). We assumed that $\lambda$ and $\xi$ have 
temperature dependence that is proportional to $(1-T/T_c)^{-1/2}$.  

We now describe in more detail the implementation of the distribution
of point defects through the sample. Following Blatter {\it et al.}
\cite{Blatter94}, we generate a random pinning energy distribution
per pancake given by
\begin{eqnarray}
  U^p(\vec{u},z)=d \int\int d^2R\
  U_{pin}(\vec{R},z)\ p(\vec{R}-\vec{u}),\nonumber \\
p(\vec{R})=2\xi^2/(R^2+2\xi^2),\\
\langle U_{pin}(\vec{R},z)U_{pin}(\vec{R'},z')\rangle=\gamma_U
\delta^{(2)}(\vec{R}-\vec{R'})\delta(z-z')\nonumber.
\end{eqnarray}
Here $\gamma_U$ gives the variance of the gaussian distributed random
variable $U_{pin}$. $\vec{R}$ and $\vec{u}$ are two dimensional
vectors in the layer labeled by $z$. Because of the convolution of the
uncorrelated 
random variable with the single pin form factor $p(\vec{R})$ we obtain
a distribution of random numbers with a power law correlation in each
layer, uncorrelated among different layers. The equations above are
discretized on a fine grid and the integrals are implemented as
summations. Typically we used a grid size of $3a_0/800$.
The value of $\gamma_U$ used in most of our simulations is
$\gamma_U(T)=\gamma_U(0) (1-T/T_c)^2$, with $\gamma_U(0)=
1.72\times 10^{-9}$erg$^2$cm$^{-3}$=$0.09$(k$_B$K)$^2$\AA$^{-3}$.
 
\begin{figure}[h]
\centering
\includegraphics[width=0.4\textwidth]{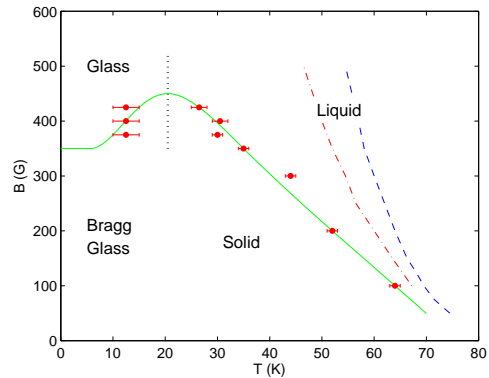}
\caption{
(color online) Phase diagram. The transition temperatures for vortices
with point disorder at various fields were obtained from simulations 
with simulated annealing and
vortex shaking ($\bullet$, red), and a smooth line (green) is drawn to fit
the data. The melting transition of the pure system (dashed, blue) is
plotted for comparison, as well as the transition line of the system
with point disorder without annealing and shaking (dashed-dot,red).
}
\label{figure1}
\end{figure}

We now discuss the results of the simulations. In Fig.\ref{figure1} we
show the phase diagram of the system with point disorder using our
technique of simulated annealing and vortex shaking. For comparison we
show the melting line of the same
system without shaking and starting from an ordered configuration of
FLs for each temperature and field. Only when using annealing and
shaking we are able to see the inverse melting for fields in the
range $350<B<450$G. The BrG phase, VG phase and
VL phase are indicated on the figure. Recent experiments
also show more subtle second order (SO) transitions between the VG and VL
phases and between the BrG and Solid phases. In the present
simulations we did not explore these SO transitions, but we indicated
their approximate location on the figure as a dotted line
as expected from the experimental results. Here we only 
measure the FO transition line. Actually the order of the transition
can only be resolved in the simulation to be of first order for small
fields $B<300$G by observing a small step in the total energy. For
higher fields, since the system is small we 
cannot resolve the difference between a weak first order and a second
order transition. 

Next, we give the details of the simulated annealing and vortex
shaking procedure. For a given magnetic field, the system is first 
simulated at a temperature
above the melting transition for 180 time units (time is measured 
\cite{Goldschmidt05,Goldschmidt07} in units  of $\eta a_0^2/\epsilon_0(T)$ 
where $\eta$ is the viscous drag coefficient per unit length and 
$\epsilon_0(T)=(\phi_0/4\pi\lambda(T))^2$). We
then decrease the temperature by steps of 0.5K and start each
subsequent simulation from the last configuration of the previous run.
For each step the simulation time of 180 units is divided as follows:
During time intervals 0-40, 80-90 and 130-140 vortex shaking is
applied as will be discussed below. After each period of shaking 20
units of time is spent for equilibration and then 20 units of time for
measurements. The vortex shaking is implemented by applying
alternating forces on all pancakes in adjacent layers. Within the same
layer the force is the same on all pancakes then in the next layer it
is opposite, and since we have an even number of layers the total
force is zero.In addition the shaking force is varying with time as
follows: For 0.5 units of time the force in the $+x$ direction, in
the next 0.5 units of time it is in the $-x$ direction, and then
similarly in the $\pm y$ direction. This complete a shaking cycle of 2
time units that can be repeated. The strength of the shaking force was
taken to be $F_{shake}=\sqrt{\pi d \gamma_U/2}$. Thus it is
proportional to the average pinning force. The annealing and shaking
procedure helps the system to reach a configuration closer to true
thermodynamic equilibrium and prevents it to be stuck in a metastable
state. 

\begin{figure}[h]
\centering
\includegraphics[width=0.5\textwidth]{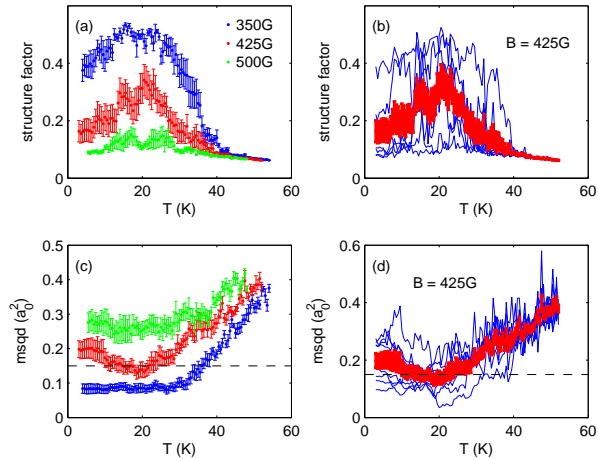}
\caption{
(color online) Structure factor (SF) and mean square deviation (msqd). 
(a) SF at 350 G (top, blue), 425 G (middle, red), and 500 G 
(bottom, green), average over 8 realizations.
(b) SF at 425 G (thick red line, average over 8 disorder
realizations), 
the results for each individual
realization are shown in the background (thin blue lines).
(c) msqd at 350 G (bottom, blue), 425 G (middle red), and 500 G 
(top, green), average over 8 realizations.
(d) msqd at 425 G, similar to (b). 
}
\label{figure2}
\end{figure}
Next we discuss the results of the measurements of the structure
factor and the mean square deviation of the FLs. These measurements
are depicted in Fig. 2. In Fig. 2(a) we show the normalized
structure factor as a function of temperature for three different
fields. The structure factor is calculated at the first Bragg peak and
is normalized to 1 for perfect crystalline order.
This figure shows the average results for 8 different
realizations of the random potential. In Figure 2(b) we show the
individual realizations for one field (425G) that make up the average.
We see that for the lower fields, as the temperature is decreased, 
the structure factor starts to rise below 40 K and reaches a maximum at
about 20 K and then decreases again. If we set the threshold at about
0.2-0.25 then we can say that for $B=425$G there is a freezing
transition at about $T=30$ K and an inverse melting transition at about
$T=10K$. For $B=350$G there is only a freezing transition at about
$T=35$K and no reentrant behavior. For $B=500$G there is no
transition at all.

In Fig.~2(c) we see the averaged mean square deviation of the FLs 
from straight
lines. In fig.~2(d) we see the eight realizations that make up the
average for $B=425$G. If we take the value of 0.15 as the borderline
case we see again that for $B=350$ there is a melting transition at
around T=35 K, for $B=425$ G there is both a freezing and an inverse
melting and for $B=500$ K there is no transition at all. Figures
2(c) and 2(d) obtained from our actual simulations are very similar to
results obtained by Ertas and Nelson \cite{Ertas96} (see Figures
3 and 4 in that reference) using the cage model together with a
transfer matrix computation of the root mean square deviation. The
corresponding $T^*$ in our case is about 10 K. The reason that the
mean square deviation first decreases as one raises the temperature
from zero is that thermal fluctuations will initially act to blur the
pinning sites and prevent the pancakes to best accommodate the random
potential.  Thus the pinning energy will be less negative and
the Josephson energy will decrease as the FLs become straighter. As the
temperature further increases it will make the FLs wiggle more because
of thermal motion and the mean square deviation increases.

\begin{figure}[h]
\centering
\includegraphics[width=0.4\textwidth]{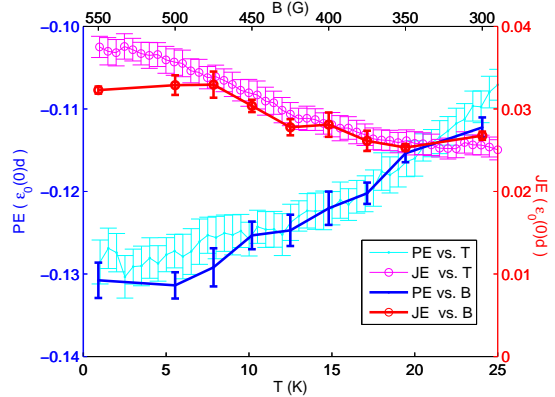}
\caption{
(color online) Pinning energy (PE) and Josephson energy (JE). The PE 
(thin, cyan) and JE (thin, magenta)
versus T curves were simulations with B = 400 G. The PE 
(thick, blue) and JE (thick, red) versus B
curves correspond to T = 15 K. 
}
\label{figure3}
\end{figure}

This argument can be further supported by observing the pinning and
Josephson energies at low temperatures, the latter reflecting the
deviation of the FLs from straight stacks. 
In Fig.~3 we observe the behavior of the
pinning energy and Josephson energy as the temperature increase for a
fixed field of 400 G. Superimposed are the same energies as a function of
decreasing field at fixed temperature of 15 K. In both cases the
system changes from the VL phase into the BrG phase. We see the
decrease of the effective pinning (pinning energy becomes less
negative, so it rises) and this is compensated by a decrease of the
Josephson energy, i.e. the FLs become straighter. This is a disorder
driven transition so the pinning energy plays a similar role to the
entropy in a 
temperature driven transition, but here the more ordered phase occurs at
higher temperature or lower magnetic field as is evident from
Fig.~1. For temperatures above 25K the Josephson energy starts to
increase again due to increased thermal fluctuations. Fig.~3 shows
that decreasing the field at fixed temperature makes the pinning less
effective and helps increase the amount of order in the system similarly
to increasing the temperature.

\begin{figure}[h]
\centering
\includegraphics[width=0.4\textwidth]{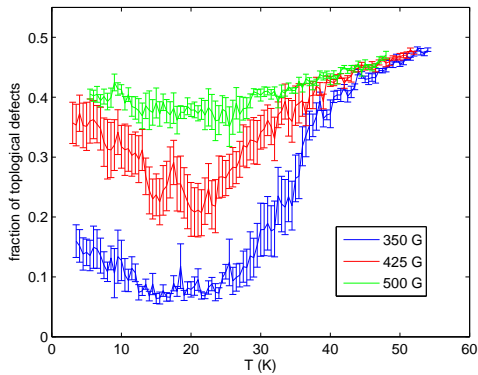}
\caption{
(color online) Fraction of topological defects. Three different fields are shown:
350 G (bottom, blue), 425 G (middle, red), and 500 G (top, green). The curve for
each field is averaged over 8 disorder realizations.
}
\label{figure4}
\end{figure}
To probe further the difference between the BrG and VG phases we
measured the fraction of topological defects (TD). For a given
configuration of pancakes we use Delaunay triangulation to measure the
number of nearest neighbors for each lattice point. Every deviation
from 6 is a topological defect. The fraction of sites with TDs to
the total number of sites, averaged over many configurations is depicted in
Fig.~4. We see this fraction for three different fields as a function
of temperature. If we impose a borderline value of 0.3 we see that for
$B=350$ G the transition to BrG at 35 K is associated with a decrease
in the number of TDs. For $B=425$ G there is an inverse
melting from BrG to VL which is associated with an
increase in the number of TDs. For $B=500$ G the fraction of TDs never
falls significantly. The abundance of TDs is also present in the
liquid so it does not distinguish the VG from a liquid. We can think
of the VG as kind of a frozen liquid where the pancakes' positions 
are frozen
near strong pinning sites. We also verified the amount of entanglement
is large in the VG similar to the liquid by measuring the number of
non-simple loops in the system.
\begin{figure}[h]
\centering
\includegraphics[width=0.45\textwidth]{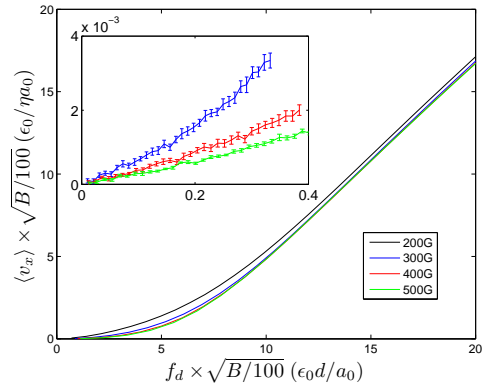}
\caption{
(color online) Drifting velocity versus driving force at 15K. 
The results for each field
are averages over 8 realizations. For the axes units are chosen such that
$a_0=a_0(100G),\ \epsilon_0=\epsilon_0(15K)$ . The insert shows a magnification of the low 
driving force region. 
}
\label{figure5}
\end{figure}
Experimentally the transition from BrG to the VG was historically
referred to as the ``second peak''. As the magnetic field is increased
at fixed temperature below $T_{sp}$, there is a jump in the critical
current at the transition. This jump is associated with an increase in
the effective pinning of the FLs, thus making them harder to detach
from their equilibrium positions by the applied Lorentz force due to
the current. We carried out measurements of the response of the
pancakes to an applied force in the $x$-direction (related to the I-V
characteristics). 
The results are depicted in Fig.~5 for different fields at $T=15$
K. We see that
for $B=300-400$ G the pancakes mobility decreases with increasing
field as 
measured by their velocity under the influence of an applied force.
For $B=450-500$ G there is no longer much change with increasing field
and the pinning seems to saturate. This seems consistent with a
transition into the VG phase.

To summarize, in this work we reproduced the inverse melting of
the vortex matter from BrG phase into the VG phase using numerical
simulations and we discussed some of the features characterizing each
phase based on quantities measured in the simulations. 

{\it Acknowledgments} - This work is supported by the US Department of
Energy (DOE), Grant No. DE-FG02-98ER45686. We
also thank the DOE NERSC program and the Pittsburgh 
Supercomputing Center for time allocations.
   
\end{document}